\documentclass{aa}
\usepackage{graphicx}
\usepackage{natbib}
\bibpunct{(}{)}{;}{a}{}{,}
\begin{document}

\title{Astrometric accuracy of phase-referenced observations with the VLBA and EVN}

\author{ N. Pradel \inst{1}
\and P. Charlot\inst{1}
\and J.-F. Lestrade \inst{2}}

\offprints{Nicolas Pradel, \email{ pradel@obs.u-bordeaux1.fr}}

\institute {Observatoire de Bordeaux (OASU)--CNRS/UMR
  5804, BP~89, 33270 Floirac, France 
\and Observatoire de Paris--CNRS/UMR 8112, 61 Avenue de
  l'Observatoire, 75014 Paris, France} 

\date{Received / Accepted }

\abstract{Astrometric accuracy of complex modern VLBI arrays cannot
be calculated analytically. We study the astrometric accuracy of
phase-referenced VLBI observations for the VLBA, EVN and global VLBI
array by simulating VLBI data for targets at declinations $-25\degr$,
$0\degr$, $25\degr$, $50\degr$, $75\degr$ and $85\degr$. The systematic
error components considered in this study are calibrator position,
station coordinate, Earth orientation and troposphere parameter
uncertainties. We provide complete tables of the astrometric
accuracies of these arrays for a source separation of $1 \degr$ either
along the right ascension axis or along the declination
axis. Astrometric accuracy is $50~\mu$as at mid declination and is
$300~\mu$as at low ($-25 \degr$) and high ($85 \degr$) declinations
for the VLBA and EVN. In extending our simulations to source
separations of $0.5 \degr$ and $2 \degr$, we establish the formula for
the astrometric accuracy of the VLBA: $\Delta_{\alpha\cos\delta ,\delta}
= (\Delta_{\alpha\cos\delta, \delta}^{1^{\circ}}-14) \times d+ 14$ ($\mu$as) where
$\Delta_{\alpha\cos\delta, \delta}^{1^{\circ}}$ is the astrometric accuracy for a separation
$d=1\degr$ provided in our tables for various declinations and
conditions of the wet troposphere. We argue that this formula is also
valid for the astrometric accuracy of the EVN and global VLBI array. 
}

\titlerunning{Astrometric accuracy of phase-referenced observations with the VLBA and EVN}
\authorrunning{Pradel, Charlot, Lestrade}
\maketitle

\section{Introduction}
\label{Introduction}

Very Long Baseline Interferometry (VLBI) narrow-angle astrometry
pioneered by Shapiro et al.~(1979) makes use of observations of pairs
of angularly close sources to cancel atmospheric phase fluctuations
between the two close lines of sight. In this initial approach, the
relative coordinates between the two strong quasars \object{3C345} and
\object{NRAO512} and other ancilliary parameters were adjusted by a
least-squares fit of the differenced phases after connecting the VLBI
phases for both sources over a multi-hour experiment. Then,
\cite{Mar83,Mar84} made the first phase-referenced map where structure
and astrometry were disentangled for the double quasar
\object{1038+528~A} and B. Both of these experiments demonstrated
formal errors at the level of a few tens of microarcseconds or
less in the relative angular separation between the two
sources. 

Another approach was designed to tackle faint target sources by
observing a strong reference source (quasar) to increase the
integration time of VLBI from a few minutes to a few hours
\citep{Les90}. This approach improves the sensitivity by the factor
$$\sqrt { N_b\times\frac{T_{int}}{T_{scan}}},$$ where $N_b$ is the
number of VLBI baselines, $T_{int}$ is the extended integration time
permitted by phase-referencing (several hours) and $T_{scan}$ is the
individual scan length (a few minutes). As this factor is very large
(e.g. $> 50$ for the 45 baselines of the Very Long Baseline Array),
faint target sources can be detected and their positions can be
concomitantly measured with high precision.

In the approach above, the VLBI phases of the strong reference
source are connected, interpolated in time and differenced with the
VLBI phases of the faint source that do not need to be connected. The
differenced visibilities are then inverted to produce the map of the
brightness distribution of the faint target source and its position is
determined by reading directly the coordinates of the map peak which
are relative to the a priori reference source coordinates. The map
is usually highly undersampled but suffices for astrometry. This  {\it
mapping astrometry} technique is implemented in the SPRINT software
\citep{Les90} and a similar procedure is also used within the NRAO
AIPS package to produce phase-referenced VLBI maps with absolute
source coordinates on the sky.

While phase-referencing in this way is efficient, it still provides no
direct positional uncertainty as does least-squares fitting of
differenced phases \citep{Sha79}. In order to circumvent this problem,
we have developed simulations to evaluate the impact of systematic
errors in the derived astrometric results. Such simulations have been
carried out for of a pair of sources observed with the Very Long
Baseline Array (VLBA) and the European VLBI Network (EVN) at various
declinations and  angular separations. Systematic errors in station
coordinates, Earth rotation parameters, reference source coordinates
and tropospheric zenith delays were studied in turn. The results of
the simulations are summarized below in tables that indicate positional
uncertainties when considering these systematic errors either
separately or altogether. Such tables can be further interpolated to
determine the accuracy of any full-track experiment with the VLBA and
EVN.

Our study includes atmospheric fluctuations caused by the turbulent
atmosphere above all stations. These fluctuations have been considered
uniform and equivalent to a delay rate noise of $0.1$~ps/s for all
stations. The impact of these fluctuations is limited if the antenna
switching cycle between the two sources is fast enough. The phase
structure function measured at 22 GHz above the VLA by \cite{Car99}
provides prescriptions on this switching time. At high frequency, it
can be as short as 10s, as e.g. in \citet{Rei03} who carried out
precise 43~GHz VLBA astrometric observations of Sgr A$^*$ at a
declination of $-28\degr$. Switching time in more clement conditions
is typically a few minutes at 8.4 GHz for northern sources. 

A few applications of {\it mapping astrometry} are the search for
extra-solar planets around radio-emitting stars \citep{Les94}, the
determination of the Gravity Probe B guide star proper motion
\citep{Leb99}, the determination of absolute motions of VLBI
components in extragalactic sources, e$.$g$.$ in compact symetric
objects \citep{Cha03} or core-jet sources \citep{Ros99}, probing the
jet collimation region in extragalactic nuclei \citep{Ly004}, pulsar
parallax and proper motion measurements \citep{Bri02} and the
determination of parallaxes and proper motions of maser sources in
the whole Galaxy as planned with the VERA project \citep{Kaw00,Hon00}.

\section{Method}
\label{Method}

As indicated in e.g. \cite{Tho86}, the theoretical precision of
astrometry with the interferometer phase is 
\begin{equation}
\sigma_{\alpha, \delta} = {1 \over {2 \pi}} ~ {1 \over SNR } 
~ { \lambda \over B} ,
\end{equation} where
$SNR$ is the signal-to-noise ratio of the observation, $\lambda$ is
the wavelength and $B$ is the baseline length projected on the sky. For
observations with the VLBA ($B\sim 8000$~km), $\lambda = 3.6$~cm, and
a modest $SNR$ of $10$, this theoretical precision is breathtakingly
$\sim 15~ \mu$as. Although a single observation of the target yields
an ambiguous position, multiple observations over several hours easily
remove ambiguities even with a sparse u-v plane coverage
\citep{Les90}.

While the theoretical precision above might be regarded as the potential
accuracy attainable for the VLBI, systematic errors in the model of
the phase limit narrow-angle astrometry precision to roughly ten times
this level in practice \citep{Fom99}. An analytical study of systematic errors
in phase-referenced VLBI astrometry over a single baseline is given in 
\citet{Sha79} and it shows that all systematic errors are scaled by
the source separation. Another error analysis in such differential VLBI
measurements can be found in \citet{Mor84}. However, for modern VLBI arrays
with 10 or more antennae, the complex geometry makes the analytical
approach intractable. For this reason, we have estimated such
systematic errors by simulating VLBI visibilities and inverting them
for a range of model parameters (station coordinates, reference source
coordinates, Earth Orientation parameters, and tropospheric dry and wet
zenith delays) corresponding to the expected errors in these parameters.

The visibilities were simulated for a pair of sources at declinations
$-25\degr$, $0\degr$, $25\degr$, $50\degr$, $75\degr$, $85\degr$ and
with angular separations $0.5\degr$, $1\degr$ and 2$\degr$ for the
VLBA, EVN and global VLBI array (VLBA+EVN). For each of these cases,
we simulated visibilities every 2.5 min from source rise to set
(full track) with a lower limit on elevation of 7$\degr$. The adopted
flux for each source (calibrator and target) was 1~Jy to make the phase
thermal noise negligeable in our simulations. For applications to faint
target sources, one should combine the corresponding thermal astrometric
uncertainty (Eq. 1) with the systematic errors derived below. The
simulated visibilities were then inverted using uniform weighting to
produce a phase-referenced map of the target source and estimate
its position. This operation was repeated 100 times in a {\it Monte
Carlo} analysis after varying slightly the parameters of the model
based on errors drawn from a Gaussian distribution with zero-mean and
plausible standard deviation. We report the rms of the differences
found between the known a priori position of the target source and the
resulting estimated positions as a measure of the corresponding systematic
errors for each of the above cases. We have adopted the usual
astrometric frequency of $8.4$~GHz for this analysis.

\section{Phase model used in simulation}
\label{model}

The phase delay and group delay in VLBI are described in
\citet{Sov98}. The phase $ \phi = \nu \tau$ at frequency $\nu $ is
related to the interferometer delay $$\tau = \tau_g + \tau_{trop}
+\tau_{iono}+ \tau_R + \tau_{struc} + \tau_{clk} .$$ Specifically, the
geometric delay is: $$\tau_g  = { { { [P][N][EOP]} { \vec b . \vec k
\over c }}} $$ with the precession matrix $[P]$, the nutation matrix
$[N]$, the Earth Orientation Parameters matrix $[EOP]$, the baseline
coordinates $\vec b$ in the terrestrial frame, the source direction
coordinates $\vec k$ computed with source right ascension and
declination in the celestial frame. The ``retarded baseline correction''
to account for Earth rotation during elapsed time $\tau_g$ must also
be modelled \citep{Sov98}. The differential tropospheric delay $\tau _{trop}$
between the two stations is computed with a static tropospheric model
and the simple mapping function $ 1/\sin E $ (where $E$ is the source
elevation at station) to transform the zenith delay into the
line-of-sight delay at each station. The differential ionospheric
phase delay $\tau_{iono} = -{{k {\rm TEC}}/ {\nu^2}}$ is  related  to
the total electronic content TEC in the direction  of the source at
each station. The General Relativity delay $\tau_R$ takes  into
account light propagation travel time in the gravitational  potential
of the Sun. The source structure contribution $\tau_{struc}$ can be
computed according to the model by \cite{Cha90} but was not included
in our simulations which are for point sources. The clock delay $\tau
_{clk}$ cancels in differenced VLBI phases.

The model above is that implemented in the SPRINT software used for
our simulations. It is thought to be complete for narrow-angle
astrometry and additional refinements, such as ocean loading, atmospheric
loading, etc\dots, would not make difference into our results. We
have not studied the ionosphere contribution to systematic errors. The
unpredictible nature of the ionosphere makes this task 
difficult. Calibration of the ionosphere by dual-frequency
observations, or over a wide bandwidth at low frequency \citep{Bri02},
or simply by observing at high frequency ($>10$~GHz) where the effect
is small, offers solutions to this problem.

\section{Results}
\label{results}

\subsection{VLBA}
\label{VLBA}

\begin{table}[t]
\caption{Adopted rms errors for the source coordinates, VLBA station
coordinates and Earth Orientation Parameters in our {\it Monte Carlo}
simulations.}
\label{errors}
\centering
\begin{tabular}{lr} \hline\hline
    Parameters                          &    Errors \\  \hline
    Source coordinates &\\
    $\alpha_0 \cos \delta_0 $           &   0.25/1 mas  \\ 
    $\delta_0$                          &   0.25/1 mas  \\    \hline
    Station coordinates&\\
    $X$                                 &   1--2  mm  \\  
    $Y$                                 &   1--3  mm  \\ 
    $Z$                                 &   1--2  mm  \\   \hline 
    Earth Orientation Parameters&\\
    $X_p$                               &   0.2 mas     \\
    $Y_p$                               &   0.2 mas    \\ 
    $UT1-UTC$                           &   0.02 ms \\
    $\psi \sin\epsilon$                 &    0.3 mas    \\
    $\epsilon$                          &    0.3 mas    \\   \hline
\end{tabular}
\end{table}
\addtolength{\tabcolsep}{-1.2mm}
\begin{table}[t]
\caption{Dry and wet tropospheric zenith path delays ($\tau_{dtrp}$
  and $\tau_{wtrp}$) at the VLBA stations along with the adopted rms
  errors $\Delta\tau_{dtrp}$ and $\Delta\tau_{wtrp}$ in our {\it Monte
  Carlo} simulations.}
\label{delay}
\centering
\begin{tabular}{lcccccccc} \hline\hline
    Stations             &\multicolumn{2}{c}{Dry trop.}&&\multicolumn{5}{c}{Wet trop.}\\
&&&&\multicolumn{2}{c}{Mean}&&\multicolumn{2}{c}{Max}\\
\noalign {\vskip 0.0mm}
\cline{2-3}\cline{5-6}\cline{8-9}
\noalign {\vskip 1.0mm}
                         &$\tau_{dtrp}$&$\Delta\tau_{dtrp}$&&$\tau_{wtrp}$&$\Delta\tau_{wtrp}$&&$\tau_{wtrp}$&$\Delta\tau_{wtrp}$\\
                         &     (cm)       &   (cm)    &&    (cm)        &   (cm)    &&   (cm)       &  (cm)  \\\hline
    Brewster             &      225       &    0.5    &&       ~8       &     2.7   &&    13        &  ~4.3  \\
    Fort Davis           &      192       &    0.5    &&       ~8       &     2.7   &&    15        &  ~5.0  \\
    Hancock              &      223       &    0.5    &&       ~9       &     3.0   &&    19        &  ~6.3  \\
    Kit Peak             &      185       &    0.5    &&       ~6       &     2.0   &&    15        &  ~5.0  \\
    Los Alamos           &      185       &    0.5    &&       ~6       &     2.0   &&    13        &  ~4.3  \\
    Mauna Kea            &      149       &    0.5    &&       ~1       &     2.0   &&    ~4        &  ~2.0  \\
    North Liberty        &      225       &    0.5    &&       10       &     3.3   &&    19        &  ~6.3  \\
    Owens Valley         &      199       &    0.5    &&       ~5       &     2.0   &&    20        &  ~6.7  \\
    Pietown              &      176       &    0.5    &&       ~4       &     2.0   &&    12        &  ~4.0  \\ 
    Saint Croix          &      213       &    0.5    &&       22       &     7.3   &&    30        &  10.0  \\ \hline
\end{tabular}
\end{table}
\addtolength{\tabcolsep}{+1.2mm}

The parameter rms errors adopted as plausible for the VLBA phase model
are listed in Tables \ref{errors} and \ref{delay}. The reference
source coordinate uncertainties ($\Delta\alpha_0\cos\delta_0$,
$\Delta\delta_0$) of 1 mas are typical of those in the VLBA Calibrator
Survey \citep{Bea02}, from which most of the reference sources
originate. However, ICRF extragalactic sources have better position
accuracies down to 0.25~mas \citep{Ma998}. We have thus carried
out the calculations for both of these cases (1~mas and 0.25~mas)
and both $\alpha_0$ and $\delta_0$ have been perturbed by these
uncertainties in our simulations. The uncertainties for the station
coordinates are from the ITRF2000 frame \citep{Bou04} while those
for the Earth Orientation Parameters are from the IERS web
site\footnote{http://hpiers.obspm.fr/iers/eop/eopc04/EOPC04.GUIDE
  (Table~2).}. The adopted dry tropospheric rms error $\Delta\tau
_{dtrp}$ of 0.5 cm corresponds to $2.5$ millibars in atmospheric
pressure uncertainty at sea level. Although barometer reading is
usually better, the absolute calibration of station barometers is at
this level. Uncertainties in the wet tropospheric zenith delay
$\tau_{wtrp}$ derived from temperature and humidity are known to be
large \citep{Saa73}. Experience makes us believe that a 30\% error is likely on
$\tau_{wtrp}$ and thus we took 1/3 of $\tau_{wtrp}$ as the plausible rms error
$\Delta\tau_{wtrp}$ with a minimum value of 2~cm. We carried out
simulations for both mean and maximum values of wet zenith path delays
based on estimates of $\tau _{wtrp}$ recently derived from multiple
VLBA geodetic and astrometric sessions \citep{Sov03}. The maximum wet
zenith delays and corresponding errors were used to investigate the
impact of extreme weather conditions on observations. These values are
listed for each VLBA station in Table \ref{delay}.

\begin{table*}[bht]
\caption{VLBA rms astrometric errors (in $\mu$as) for a relative source
separation $(\alpha -\alpha _0)\cos\delta_0 = 1\degr$. Individual astrometric error contributions
from calibrator position, Earth orientation parameter, station coordinate, and 
dry and wet troposphere uncertainties are given separately, while the last
two lines indicate the total astrometric errors when all model parameters
are perturbed together.}
\label{VLBARA}
\centering
\vskip 1.5mm \small \tabcolsep=0.82mm
\begin{tabular}{lccccccccccccccccccc}
\hline
\hline
\noalign {\vskip 1.0mm}
&&\multicolumn{16}{c}{Declination of source}\\
\noalign {\vskip 0.5mm} \cline{3-19} \noalign {\vskip 1.0mm} \hfil
\hfil
&&\multicolumn{2}{c}{$-25\degr$}&&\multicolumn{2}{c}{$0\degr$}&&\multicolumn{2}{c}{$25\degr$}
        &&\multicolumn{2}{c}{$50\degr$}&&\multicolumn{2}{c}{$75\degr$}&&\multicolumn{2}{c}{$85\degr$}\\
\noalign {\vskip 0.0mm}
\cline{3-4}\cline{6-7}\cline{9-10}\cline{12-13}\cline{15-16}\cline{18-19}
\noalign {\vskip 1.0mm}
Error component      &&$\Delta\alpha \cos\delta \hfil$ &$\Delta\delta$ &&$\Delta\alpha \cos\delta \hfil$ &$\Delta\delta$&&
        $\Delta\alpha \cos\delta \hfil$&$\Delta\delta$&&$\Delta\alpha \cos\delta \hfil$&$\Delta\delta$&&
        $\Delta\alpha \cos\delta \hfil$&$\Delta\delta$&&$\Delta\alpha \cos\delta \hfil$&$\Delta\delta$\\
\noalign {\vskip 0.0mm}
\hline
Calibrator position (1~mas error)&&   ~8   &   ~~7   &&   ~1   &   ~9 && ~8   &   16  &&   20  &   26  &&   ~59  &   ~68 && 196 & 193 \\
Calibrator position (0.25~mas error)&&~2   &   ~~7   &&   ~1   &   ~3 && ~2   &   ~5  &&   ~3  &   ~5  &&   ~12  &   ~11 && ~49 & ~50 \\
Earth orientation               &&   ~1   &   ~~8   &&   ~1   &   ~5 && ~1   &   ~6  &&   ~1  &   ~5  &&   ~~1  &   ~~4 && ~~1 & ~~4 \\
Antenna position                &&   ~2   &   ~~8   &&   ~2   &   ~4 && ~2   &   ~4  &&   ~2  &   ~3  &&   ~~2  &   ~~3 && ~~2 & ~~3 \\
Dry troposphere                 &&   15   &   ~45   &&   ~9   &   16 && ~7   &   ~9  &&   10  &   11  &&   ~18  &   ~23 && ~14 & ~16 \\
Wet troposphere (mean)          &&   53   &   182   &&   34   &   57 && 33   &   28  &&   31  &   45  &&   ~54  &   ~72 && ~79 & ~88 \\
Wet troposphere (max)           &&   87   &   219   &&   46   &   66 && 42   &   38  &&   49  &   56  &&   ~65  &   ~78 && ~81 & ~91 \\
\hline
Total (mean wtrp)               &&   60   &   175   &&   36   &   50 && 33   &   32  &&   37  &   53  &&   ~87  &   103 && 227 & 258 \\
Total (max wtrp)                &&   85   &   217   &&   43   &   74 && 42   &   44  &&   46  &   66  &&   100  &   117 && 226 & 240 \\
\hline
\end{tabular}
\end{table*}
\begin{table*}[bht]
\caption{VLBA rms astrometric errors (in $\mu$as) for a relative source
separation $\delta -\delta _0= 1\degr$. Individual astrometric error contributions
from calibrator position, Earth orientation parameter, station coordinate, and 
dry and wet troposphere uncertainties are given separately, while the last
two lines indicate the total astrometric errors when all model parameters
are perturbed together.}
\label{VLBADC}
\centering
\vskip 1.5mm \small \tabcolsep=0.82mm
\begin{tabular}{lccccccccccccccccccc}
\hline
\hline
\noalign {\vskip 1.0mm}
 &&\multicolumn{16}{c}{Declination of source}\\
\noalign {\vskip 0.5mm} \cline{3-19} \noalign {\vskip 1.0mm} \hfil
\hfil
&&\multicolumn{2}{c}{$-25\degr$}&&\multicolumn{2}{c}{$0\degr$}&&\multicolumn{2}{c}{$25\degr$}
        &&\multicolumn{2}{c}{$50\degr$}&&\multicolumn{2}{c}{$75\degr$}&&\multicolumn{2}{c}{$85\degr$}\\
\noalign {\vskip 0.0mm}
\cline{3-4}\cline{6-7}\cline{9-10}\cline{12-13}\cline{15-16}\cline{18-19}
\noalign {\vskip 1.0mm}
Error component     &&$\Delta\alpha\cos\delta \hfil$ &$\Delta\delta$ &&$\Delta\alpha\cos\delta \hfil$ &$\Delta\delta$&&
        $\Delta\alpha\cos\delta \hfil$&$\Delta\delta$&&$\Delta\alpha\cos\delta \hfil$&$\Delta\delta$&&
        $\Delta\alpha\cos\delta \hfil$&$\Delta\delta$&&$\Delta\alpha\cos\delta \hfil$&$\Delta\delta$\\
\noalign {\vskip 0.0mm}
\hline
Calibrator position (1~mas error)   &&   ~~7   &   ~~8   &&   ~1  &   ~~7 && ~8  &   11  &&   21  &   ~2  &&   59  &   ~2 && 199 & ~1 \\
Calibrator position (0.25~mas error) &&   ~~2   &   ~~7   &&   ~1  &   ~~3 && ~2  &   ~4  &&   ~5  &   ~2  &&   20  &   ~1 && ~43 & ~1\\
Earth orientation               &&   ~~5   &   ~~7   &&   ~5  &   ~~3 && ~5  &   ~4  &&   ~4  &   ~3  &&   ~3  &   ~1 && ~~3 & ~1 \\
Antenna position                &&   ~~2   &   ~~9   &&   ~2  &   ~~6 && ~2  &   ~5  &&   ~3  &   ~3  &&   ~2  &   ~2 && ~~2 & ~2 \\
Dry troposphere                 &&   ~17   &   ~54   &&   ~6  &   ~19 && ~2  &   12  &&   ~5  &   ~9  &&   11  &   ~9 && ~12 & 13 \\
Wet troposphere (mean)          &&   ~80   &   272   &&   32  &   ~98 && 11  &   41  &&   12  &   32  &&   43  &   41 && ~60 & 62 \\
Wet troposphere (max)           &&   112   &   358   &&   43  &   114 && 19  &   61  &&   17  &   46  &&   59  &   55 && ~74 & 71 \\
\hline
Total (mean wtrp)                &&   ~84   &   284   &&   30  &   ~99 && 16  &   42  &&   25  &   33  &&   81  &   36 && 189 & 67 \\
Total (max wtrp)                 &&   121   &   481   &&   44  &   134 && 20  &   56  &&   26  &   46  &&   92  &   65 && 212 & 74 \\
\hline
\end{tabular}
\end{table*}

We simulated the visibilities of a full u-v track experiment with the
VLBA for six declinations between $-25\degr$ and $85\degr$
with a $1\degr$ relative source separation (either oriented in
right ascension or in declination). Uniform weighting was applied
to the visibilities, resulting in a synthesized beam mainly shaped by the
longest baselines. As a test, we have also removed the 9 baselines
smaller than 1500~km in length out of the 45 baseline array and
noted a decrease in systematic errors of $\sim 15\%
$ in a few test cases. Conservatively, we have retained these ``short''
baselines in our final simulations. This is motivated by the fact that all
possible baselines must be kept for sensitivity when observing weak sources.
The antenna switching cycle
between target and reference sources was set to 2.5 minutes. The results,
however, do not depend critically on this value. It was chosen so
that the automatic phase connection routine for the reference source
does not discard too much data in the presence of a delay rate error of
0.1~ps/s (adopted uniformly for all the stations in the
simulation). As mentioned previously, we analysed these data simulated
with SPRINT using the {\it a priori} parameter values perturbed by some
errors. We carried out this analysis 100~times for each systematic error
component with perturbation errors drawn from Gaussian distributions
with zero mean and standard deviations according to the rms errors in
Tables~\ref{errors} and \ref{delay}. The resulting position of the
target was estimated by reading the peak position in each of the 100 
phase-referenced maps. The pixel size in the maps was
0.05~mas. This size is small compared to the synthesized beam ($\sim1$~mas
at 8.4~GHz on 8000~km baseline) and, hence, the uncertainty in the peak
position due to the pixel size is negligeable. This position was
determined by fitting a parabola over the full half beam width. This
procedure was used in the Hipparcos/VLBI work of \cite{Les99} and was
found to be appropriate. As expected, each position was slightly
offset from the map phase center, reflecting the corresponding
systematic errors. After substracting the initial perturbation in
the calibrator position, we calculated the rms of these 100 relative
coordinate offsets $\Delta\alpha\cos\delta$ and $\Delta\delta$ for the adopted
$1\degr$ source separation in right ascension or declination. Note
that the mean of these 100 coordinate offsets was close to zero in all
cases. In Tables \ref{VLBARA} and \ref{VLBADC}, we report the rms
astrometric errors for each individual error component along with
the total astrometric errors when all model parameters are perturbed
together in the simulation. The total errors were derived
by considering a 1~mas error in the calibration position.

The wet troposphere systematic error clearly dominates over all the
other error components for $\delta \leq 50 \degr$ but the calibrator
error dominates at higher declinations if its position is not known
to better than 1~mas. 
This behavior was first noted by \citet{Sha79} who derived analytical
formulae providing the astrometric errors caused by the calibrator coordinate
uncertainties in the case of a single VLBI baseline. A detailed analysis
comparing our simulated errors with those obtained from these formulae
is given in Appendix A. Other systematic errors,
in particular the Earth orientation parameter and the station
coordinate errors, are small. In Tables~\ref{VLBARA} and~\ref{VLBADC},
we note that astrometric errors originating from mean and maximum wet
troposphere uncertainties are not drastically different (a ratio of 1.5
at most).

\begin{figure*}[htb]
\centering
\includegraphics[height=12cm, width=8cm, angle=-90]{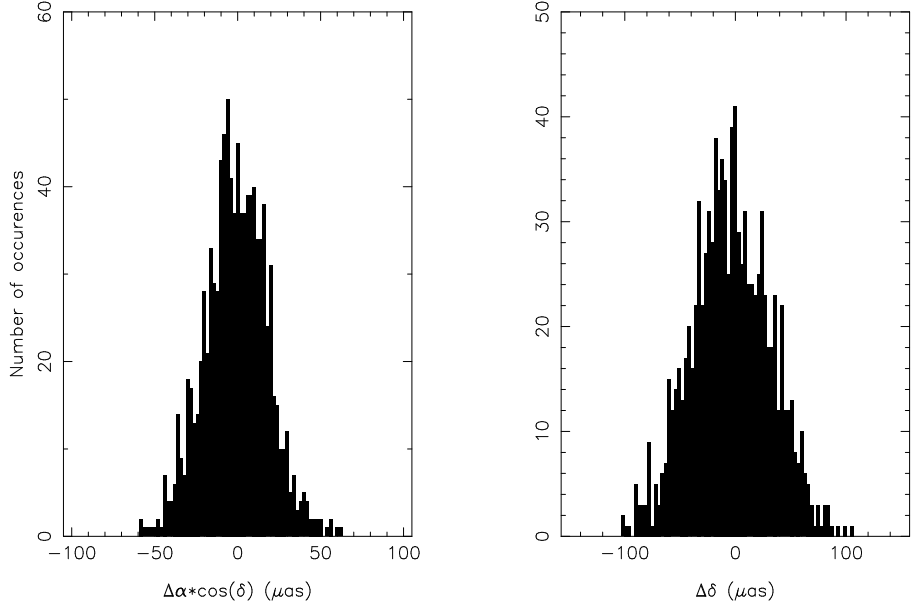}
\caption{Distribution of total astrometric errors for a $1\degr$
  relative source separation along declination at declination
  $50\degr$. All perturbating errors (calibrator position, Earth
  orientation parameters, station coordinates, dry and wet troposphere
  parameters) are considered together in this simulation.}
\label{chiDC}
\end{figure*}

Finally, we have plotted in Fig$.$ \ref{chiDC} the distribution of all
coordinate offsets $\Delta \alpha \cos \delta$ and $\Delta \delta$ for
the $50\degr$ declination target when all perturbation errors are 
present. For this specific case we have carried out 1000 simulations to
refine the binning of the distribution. We have also performed the
Pearson test on all distributions and provide the reduced chi-square
$\chi^2_{\nu}$ and probability $p$ that such distributions are Gaussian
in Table~\ref{chi}. The results of this test show that most of the
distributions are not Gaussian with $p$ generally smaller than
$0.4$.

\begin{table}[t]
\centering
\caption{Reduced chi-square $\chi ^2_\nu$ and probability $p$ of Gaussian distribution 
for the astrometric errors $\Delta \alpha \cos \delta$ and $\Delta \delta$ using the Pearson test.}
\label{chi}
\begin{tabular}{l|cc|cc} \hline\hline
               &\multicolumn{4}{c}{$(\alpha -\alpha _0)\cos\delta_0 = 1^{\circ}$}\\
Declination    &\multicolumn{2}{c}{$\Delta \alpha \cos \delta$}&\multicolumn{2}{c}{$\Delta \delta$}  \\ \hline
               &  $\chi ^2_{\nu}$  &  $p$    &  $\chi ^2_{\nu}$  &  $p$   \\ \hline
$-25 \degr$   &  1.51  & 1.43$\times 10^{-2}$&  2.10  & 9.59$\times 10^{-6}$\\
$0 \degr$     &  1.67  & 1.29$\times 10^{-2}$&  1.73  & 1.95$\times 10^{-3}$\\
$ 25 \degr$   &  2.33  & 8.12$\times 10^{-8}$&  1.97  & 1.12$\times 10^{-3}$\\
$ 50 \degr$   &  0.97  & 0.511  &  1.18  & 0.309 \\
$ 75 \degr$   &  0.84  & 0.783  &  1.22  & 0.273 \\
$ 85 \degr$   &  1.06  & 0.659  &  1.02  & 0.424 \\\hline
\noalign {\vskip 1.5mm}
               &\multicolumn{4}{c}{$(\delta -\delta _0)= 1^{\circ}$}\\
Declination    &\multicolumn{2}{c}{$\Delta \alpha \cos \delta$}&\multicolumn{2}{c}{$\Delta \delta$}  \\ \hline
              & $\chi ^2_{\nu}$  &  $p$    &  $\chi ^2_{\nu}$  &  $p$   \\ \hline
$-25 \degr$   &  1.19  & 0.286  &  0.76  & 0.885 \\
$0 \degr$     &  1.62  & 2.24$\times 10^{-2}$&  1.35  & 6.01$\times 10^{-2}$\\
$ 25 \degr$   &  3.22  & 4.62$\times 10^{-4}$&  1.91  & 5.57$\times 10^{-4}$\\
$ 50 \degr$   &  2.23  & 5.19$\times 10^{-6}$&  2.18  & 1.28$\times 10^{-4}$\\
$ 75 \degr$   &  1.38  & 0.116  &  2.05  & 1.99$\times 10^{-3}$\\
$ 85 \degr$   &  1.22  & 0.260  &  2.98  & 5.62$\times 10^{-12}$\\\hline
\end{tabular}
\end{table}

\subsection{EVN}
\label{EVN}

We have carried out a similar study for the EVN by simulating full
track observations for the 10 stations of the array at $8.4$~GHz. The
adopted errors for the reference source coordinates and Earth
orientation parameters were identical  to those used in the VLBA
simulations. Station coordinate errors were similar to the VLBA ones
(1--6~mm), with the exception of those for Westerbork which are at the
level of 50 mm \citep{Cha02}. The same scheme as that
adopted for the VLBA was used to define zenith dry and wet
tropospheric delay errors at each EVN station and the corresponding
values are given in Table \ref{EVNdelay}.

\addtolength{\tabcolsep}{-1.3mm}
\begin{table}[t]
\centering
\caption{Dry and wet tropospheric zenith path delays ($\tau_{dtrp}$
  and $\tau_{wtrp}$) at the EVN stations along with the adopted rms
  errors $\Delta\tau_{dtrp}$ and $\Delta\tau_{wtrp}$ in our {\it Monte
  Carlo} simulations.}
\label{EVNdelay}
\centering
\begin{tabular}{lcccccccc} \hline\hline
    Stations             &\multicolumn{2}{c}{Dry trop.}&&\multicolumn{5}{c}{Wet trop.}\\
&&&&\multicolumn{2}{c}{Mean}&&\multicolumn{2}{c}{Max}\\
\noalign {\vskip 0.0mm}
\cline{2-3}\cline{5-6}\cline{8-9}
\noalign {\vskip 1.0mm}
                         &$\tau_{dtrp}$&$\Delta\tau_{dtrp}$&&$\tau_{wtrp}$&$\Delta\tau_{wtrp}$&&$\tau_{wtrp}$&$\Delta\tau_{wtrp}$\\
                         &     (cm)       &   (cm)    &&    (cm)        &   (cm)    &&   (cm)       &  (cm)  \\\hline
    Effelsberg           &      220       &    0.5    &&       ~8       &     ~2.7   &&    20         &  ~6.7  \\     
    Hartebeesthoek       &      199       &    0.5    &&       10       &     ~3.3   &&    17         &  ~5.7  \\     
    Medicina             &      231       &    0.5    &&       11       &     ~3.7   &&    18         &  ~6.0  \\     
    Noto                 &      229       &    0.5    &&       12       &     ~4.0   &&    20         &  ~6.7  \\     
    Onsala               &      230       &    0.5    &&       ~8       &     ~2.7   &&    14         &  ~4.7  \\     
    Sheshan              &      231       &    0.5    &&       22       &     ~7.3   &&    36         &  12.0  \\     
    Urumqi               &      210       &    0.5    &&       10       &     ~3.3   &&    10         &  ~3.3  \\     
    Westerbork           &      220       &    0.5    &&       ~8       &     ~2.7   &&    20         &  ~6.7  \\      
    Wettzell             &      215       &    0.5    &&       ~7       &     ~2.3   &&    13         &  ~4.3  \\    
    Yebes                &      208       &    0.5    &&       ~5       &     ~2.0   &&    ~5         &  ~2.0  \\     \hline
\end{tabular}
\end{table}
\addtolength{\tabcolsep}{+1.3mm}

\begin{table*}[htb]
\centering
\caption{EVN rms astrometric errors (in $\mu$as) for a relative source separation
$(\alpha -\alpha _0)\cos\delta_0 = 1\degr$. Individual astrometric error contributions from station
coordinate and wet troposphere uncertainties are given separately along with the
total astrometric errors when all model parameters are perturbed together.}
\label{EVNRA}
\vskip 1.5mm \small \tabcolsep=1.32mm
\begin{tabular}{lcccccccccccccccc}
\hline \hline
\noalign {\vskip 1.0mm}
 &&\multicolumn{13}{c}{Declination of source}\\
\noalign {\vskip 0.5mm} \cline{3-16} \noalign {\vskip 1.0mm} \hfil
\hfil
&&\multicolumn{2}{c}{$0\degr$}&&\multicolumn{2}{c}{$25\degr$}
        &&\multicolumn{2}{c}{$50\degr$}&&\multicolumn{2}{c}{$75\degr$}&&\multicolumn{2}{c}{$85\degr$}\\
\noalign {\vskip 0.0mm}
\cline{3-4}\cline{6-7}\cline{9-10}\cline{12-13}\cline{15-16}
\noalign {\vskip 1.0mm}
Error component&&$\Delta\alpha\cos\delta \hfil$ &$\Delta\delta$&&
        $\Delta\alpha\cos\delta \hfil$&$\Delta\delta$&&$\Delta\alpha\cos\delta \hfil$&$\Delta\delta$&&
        $\Delta\alpha\cos\delta \hfil$&$\Delta\delta$&&$\Delta\alpha\cos\delta \hfil$&$\Delta\delta$\\
\noalign {\vskip 0.0mm}
\hline
Antenna position                &&   ~5   &   ~4 && ~5   &   ~4  &&   ~6  &   ~5  &&   ~~5  &   ~5 && ~~5 & ~~5 \\
Wet troposphere (mean)          &&   55   &   11 && 37   &   14  &&   52  &   33  &&   ~73  &   40 && ~65 & ~31 \\
\hline
Total (mean wtrp)               &&   57   &   12 && 44   &   15  &&   57  &   45  &&   ~91  &   81 && 206 & 185 \\
\hline
\end{tabular}
\end{table*}
\begin{table*}[htb]
\centering
\caption{EVN rms astrometric errors (in $\mu$as) for a relative source separation
$\delta -\delta _0= 1\degr$. Individual astrometric error contributions from station
coordinate and wet troposphere uncertainties are given separately along with the
total astrometric errors when all model parameters are perturbed together.} 
\label{EVNDC}
\vskip 1.5mm \small \tabcolsep=1.32mm
\begin{tabular}{lcccccccccccccccc}
\hline \hline
\noalign {\vskip 1.0mm}
 &&\multicolumn{13}{c}{Declination of source}\\
\noalign {\vskip 0.5mm} \cline{3-16} \noalign {\vskip 1.0mm} \hfil
\hfil
&&\multicolumn{2}{c}{$0\degr$}&&\multicolumn{2}{c}{$25\degr$}
        &&\multicolumn{2}{c}{$50\degr$}&&\multicolumn{2}{c}{$75\degr$}&&\multicolumn{2}{c}{$85\degr$}\\
\noalign {\vskip 0.0mm}
\cline{3-4}\cline{6-7}\cline{9-10}\cline{12-13}\cline{15-16}
\noalign {\vskip 1.0mm}
Error component &&$\Delta\alpha\cos\delta \hfil$ &$\Delta\delta$&&
        $\Delta\alpha\cos\delta \hfil$&$\Delta\delta$&&$\Delta\alpha\cos\delta \hfil$&$\Delta\delta$&&
        $\Delta\alpha\cos\delta \hfil$&$\Delta\delta$&&$\Delta\alpha\cos\delta \hfil$&$\Delta\delta$\\

\noalign {\vskip 0.0mm}
\hline
Antenna position                &&   ~7  &   ~6 && ~7  &   ~5  &&   ~4  &   ~~5  &&   ~5  &   ~5 && ~~4 & ~5 \\
Wet troposphere (mean)          &&   51  &   31 && 29  &   54  &&   18  &   81  &&   31  &   58 && ~33 & 68 \\
\hline
Total (mean wtrp)                &&   62  &   29 && 33  &   57  &&   27  &   78  &&   79  &   61 && 201 & 61 \\
\hline
\end{tabular}
\end{table*}

Since the EVN comprises antennas with different sensitivities, each
baseline has been weighted by the reciprocal of their noise
power equivalent $\sqrt{{\rm SEFD}_1\times {\rm SEFD}_2}$ with System
Equivalent Flux Densities (SEFD$_i$) for each station according to
Table 2 of the EVN Status
Table\footnote{http://www.mpifr-bonn.mpg.de/EVN/EVNstatus.txt} (as
available in May 2003). The Effelsberg--Westerbork baseline is 
the most sensitive baseline of the array but also the shortest one and
so unfavorable for high-accuracy astrometry. For this reason, we
decided to perform the simulations without this baseline, hence using
an array of 44 baselines only. We have applied uniform weighting to
the visibilities similarly to the VLBA. We have tested that in removing
the 12 baselines shorter than 1500~km in this 44 baseline array,
systematic errors decrease by $\sim 20\%
$ but, conservatively, we have kept them
in our simulations. In order to reduce the number of simulations,
calculations were carried out for only mean values of the wet zenith
tropospheric delays since the results when using mean or maximum
values were not found to be drasticaly different. We also did not
calculate individual contributions from calibrator
position, dry tropospheric zenith delay and Earth orientation parameter
errors since these were found to be very small for the VLBA (see
Tables \ref{VLBARA} and \ref{VLBADC}). One should keep in mind, however,
that calibrator error dominates at high declination. The results
of the EVN simulations are reported in Tables \ref{EVNRA} and \ref{EVNDC}
for a $1\degr$ source separation in right ascension or declination.

At declination $-25\degr$, many SPRINT maps were found to be
ambiguous, i.e. the main lobe of the point spread function of the EVN
could not be identified because secondary lobes were too
high. This is essentially caused by the relatively high latitude of
the array and hence to the difficulty of observing such low declination
sources due to very limited visibility periods. For this reason, we do
not provide EVN results for this declination. For other declinations,
EVN astrometric errors (Tables \ref{EVNRA} and \ref{EVNDC}) are similar
to those found for the VLBA (Tables \ref{VLBARA} and \ref{VLBADC}) and
the Westerbork position error is not a limiting factor. Declination 
accuracies are somewhat better for the EVN than
for the VLBA at low declination ($0\degr$ and $25\degr$), a consequence of
the participation of Hartebeeshoek (South Africa) in such observations.

\subsection{Global VLBI array}

\begin{table*}[hbt]
\centering
\caption{Global VLBI array rms astrometric errors (in $\mu$as) for a relative source separation
$(\alpha -\alpha _0)\cos\delta_0 = 1\degr$. The individual astrometric error contribution from wet troposphere
uncertainties is given separately along with the total astrometric errors when all
model parameters are perturbed together.}
\label{GlobalRA}
\vskip 1.5mm \small \tabcolsep=1.32mm
\begin{tabular}{lccccccccccccccccccc}
\hline\hline
\noalign {\vskip 1.0mm}
&&\multicolumn{16}{c}{Declination of source}\\
\noalign {\vskip 0.5mm} \cline{3-19} \noalign {\vskip 1.0mm} \hfil
\hfil
&&\multicolumn{2}{c}{$-25\degr$}&&\multicolumn{2}{c}{$0\degr$}&&\multicolumn{2}{c}{$25\degr$}
        &&\multicolumn{2}{c}{$50\degr$}&&\multicolumn{2}{c}{$75\degr$}&&\multicolumn{2}{c}{$85\degr$}\\
\noalign {\vskip 0.0mm}
\cline{3-4}\cline{6-7}\cline{9-10}\cline{12-13}\cline{15-16}\cline{18-19}
\noalign {\vskip 1.0mm}
Error component &&$\Delta\alpha\cos\delta \hfil$ &$\Delta\delta$ &&$\Delta\alpha\cos\delta \hfil$ &$\Delta\delta$&&
        $\Delta\alpha\cos\delta \hfil$&$\Delta\delta$&&$\Delta\alpha\cos\delta \hfil$&$\Delta\delta$&&
        $\Delta\alpha\cos\delta \hfil$&$\Delta\delta$&&$\Delta\alpha\cos\delta \hfil$&$\Delta\delta$\\
\noalign {\vskip 0.0mm}
\hline
Wet troposphere (mean)   &&   71  &   76   &&   32   &   42 && 26   &   34  &&   23  &   13  &&   22  &   ~6 && ~27 & ~~9 \\
\hline
Total (mean wtrp)       &&   82  &   67   &&   34   &   46 && 24   &   44  &&   34  &   33  &&   64  &   76 && 196 & 203\\
\hline
\end{tabular}
\end{table*}
\begin{table*}[hbt]
\centering
\caption{Global VLBI array rms astrometric errors (in $\mu$as) for a relative source separation
$\delta -\delta _0= 1\degr$. The individual astrometric error contribution from wet troposphere
uncertainties is given separately along with the total astrometric errors when all
model parameters are perturbed together.}
\label{GlobalDC}
\vskip 1.5mm \small \tabcolsep=1.32mm
\begin{tabular}{lccccccccccccccccccc}
\hline\hline
\noalign {\vskip 1.0mm}
&&\multicolumn{16}{c}{Declination of source}\\
\noalign {\vskip 0.5mm} \cline{3-19} \noalign {\vskip 1.0mm} \hfil
\hfil
&&\multicolumn{2}{c}{$-25\degr$}&&\multicolumn{2}{c}{$0\degr$}&&\multicolumn{2}{c}{$25\degr$}
        &&\multicolumn{2}{c}{$50\degr$}&&\multicolumn{2}{c}{$75\degr$}&&\multicolumn{2}{c}{$85\degr$}\\
\noalign {\vskip 0.0mm}
\cline{3-4}\cline{6-7}\cline{9-10}\cline{12-13}\cline{15-16}\cline{18-19}
\noalign {\vskip 1.0mm}
Error component &&$\Delta\alpha\cos\delta \hfil$ &$\Delta\delta$ &&$\Delta\alpha\cos\delta \hfil$ &$\Delta\delta$&&
        $\Delta\alpha\cos\delta \hfil$&$\Delta\delta$&&$\Delta\alpha\cos\delta \hfil$&$\Delta\delta$&&
        $\Delta\alpha\cos\delta \hfil$&$\Delta\delta$&&$\Delta\alpha\cos\delta \hfil$&$\Delta\delta$\\
\noalign {\vskip 0.0mm}
\hline
Wet troposphere (mean)   &&  60   &   305   &&   26  &   71 && 10  &   43  &&   ~9  &   44  &&   ~5  &   21 && ~~5 & 26 \\
\hline
Total (mean wtrp)        &&  61   &   279   &&   24  &   78 && 15  &   45  &&   22  &   46  &&   61  &   24 && 183 & 27 \\
\hline
\end{tabular}
\end{table*}

We have carried out a similar study for the global VLBI array which is the
combination of the VLBA and EVN. It includes 20 stations, with 190
possible baselines. As discussed above, the Effelsberg--Westerbork
baseline was ignored and the calculations were thus carried out for
189~baselines only. The adopted systematic error values for the
simulations with this array were the same as those adopted for the
individual VLBA and EVN (Tables \ref{errors}, \ref{delay} and
\ref{EVNdelay}) and calculations were performed for full track
observations as previously. The results of these simulations (Tables
\ref{GlobalRA} and \ref{GlobalDC}) indicate that the  astrometric
errors for the global VLBI array are consistent with those found for
the VLBA and the EVN. As expected, these errors are generally
slightly better than the ones derived for each individual array.

\section{Discussion }
\label{discussion}

\subsection{General results}

Our simulations show that the astrometric accuracy of the VLBI
phase-referencing technique (defined as
$\sqrt{(\Delta\alpha\cos\delta)^2+(\Delta\delta)^2}$) is $\sim
50~\mu$as for mid declinations and is $\leq 300~\mu$as at low and high 
declinations for point sources with a relative separation of
$1\degr$. The major systematic error components are the wet tropospheric
delay and the calibrator astrometric position, the latter only at
high declination. Station coordinate, Earth orientation parameter and
dry tropospheric zenith delay errors contribute generally to less than
$20~\mu$as in the error budget.

\subsection{Simulation of the VLBA without Saint Croix} 

\begin{table*}[hbt]
\centering
\caption{VLBA without Saint Croix rms astrometric errors (in $\mu$as) for a relative source separation
$(\alpha -\alpha _0)\cos\delta_0 = 1\degr$. The individual astrometric error contribution from wet troposphere
uncertainties is given separately along with the total astrometric errors when all
model parameters are perturbed together.} 
\label{SCRA}
\vskip 1.5mm \small \tabcolsep=1.32mm
\begin{tabular}{lccccccccccccccccccc}
\hline\hline
\noalign {\vskip 1.0mm}
 &&\multicolumn{16}{c}{Declination of source}\\
\noalign {\vskip 0.5mm} \cline{3-19} \noalign {\vskip 1.0mm} \hfil
\hfil
&&\multicolumn{2}{c}{$-25\degr$}&&\multicolumn{2}{c}{$0\degr$}&&\multicolumn{2}{c}{$25\degr$}
        &&\multicolumn{2}{c}{$50\degr$}&&\multicolumn{2}{c}{$75\degr$}&&\multicolumn{2}{c}{$85\degr$}\\
\noalign {\vskip 0.0mm}
\cline{3-4}\cline{6-7}\cline{9-10}\cline{12-13}\cline{15-16}\cline{18-19}
\noalign {\vskip 1.0mm}
Error component &&$\Delta\alpha\cos\delta \hfil$ &$\Delta\delta$ &&$\Delta\alpha\cos\delta \hfil$ &$\Delta\delta$&&
        $\Delta\alpha\cos\delta \hfil$&$\Delta\delta$&&$\Delta\alpha\cos\delta \hfil$&$\Delta\delta$&&
        $\Delta\alpha\cos\delta \hfil$&$\Delta\delta$&&$\Delta\alpha\cos\delta \hfil$&$\Delta\delta$\\
\noalign {\vskip 0.0mm}
\hline
Wet troposphere (mean)          &&  62 & 171&&       28& 47&&        27& 57&&       34& 65&&         46& 63&&       ~37& ~67\\
\hline
Total (mean wtrp)               &&63 & 193&&       27& 41&&        31& 62&&       42& 68&&         83& 82&&       211& 207\\
\hline
\end{tabular}
\end{table*}
\begin{table*}[hbt]
\centering
\caption{VLBA without Saint Croix rms astrometric errors (in $\mu$as) for a relative source separation
$\delta -\delta _0= 1\degr$. The individual astrometric error contribution from wet troposphere
uncertainties is given separately along with the total astrometric errors when all
model parameters are perturbed together.} 
\label{SCDC}
\vskip 1.5mm \small \tabcolsep=1.32mm
\begin{tabular}{lccccccccccccccccccc}
\hline\hline
\noalign {\vskip 1.0mm}
 &&\multicolumn{16}{c}{Declination of source}\\
\noalign {\vskip 0.5mm} \cline{3-19} \noalign {\vskip 1.0mm} \hfil
\hfil
&&\multicolumn{2}{c}{$-25\degr$}&&\multicolumn{2}{c}{$0\degr$}&&\multicolumn{2}{c}{$25\degr$}
        &&\multicolumn{2}{c}{$50\degr$}&&\multicolumn{2}{c}{$75\degr$}&&\multicolumn{2}{c}{$85\degr$}\\
\noalign {\vskip 0.0mm}
\cline{3-4}\cline{6-7}\cline{9-10}\cline{12-13}\cline{15-16}\cline{18-19}
\noalign {\vskip 1.0mm}
Error component  &&$\Delta\alpha\cos\delta \hfil$ &$\Delta\delta$ &&$\Delta\alpha\cos\delta \hfil$ &$\Delta\delta$&&
        $\Delta\alpha\cos\delta \hfil$&$\Delta\delta$&&$\Delta\alpha\cos\delta \hfil$&$\Delta\delta$&&
        $\Delta\alpha\cos\delta \hfil$&$\Delta\delta$&&$\Delta\alpha\cos\delta \hfil$&$\Delta\delta$\\
\noalign {\vskip 0.0mm}
\hline
Wet troposphere (mean)          &&183 & 563&&       62& 189&&        19& 55&&       17& 41&&         39& 27&&         ~42& 44\\
\hline
Total (mean wtrp)                &&190 & 534&&       70& 191&&        24& 71&&       26& 42&&         74& 28&&         216& 40\\
\hline
\end{tabular}
\end{table*}

We speculated that if the VLBA station at Saint Croix in the Virgin
Islands that suffers from dampness were withdrawn from the array, it should
improve the astrometric accuracy of the VLBA. We thus repeated our
VLBA simulations without that station. The results of this test are given
in Tables \ref{SCRA} and \ref{SCDC}. In contrast to our intuition,
the astrometric accuracy is actually degraded when the target-calibrator
direction is oriented along declination. In fact, the addition of Saint
Croix strengthens the geometry of the array and improves the astrometric
accuracy despite severe weather
conditions. In order to further explore this question, we ran simulations
without Pie Town in the middle of the array and without Mauna Kea at
the far West of the array. Withdrawing Pie Town does not change the
astrometric accuracy but the absence of Mauna Kea degrades the accuracy
in a similar way to Saint Croix.

\subsection{Linearity of the astrometric accuracy with source separation}

\begin{figure*}[htb]
\centering
\includegraphics[width=8cm,totalheight=8cm,angle=90, origin=c,angle=-90]{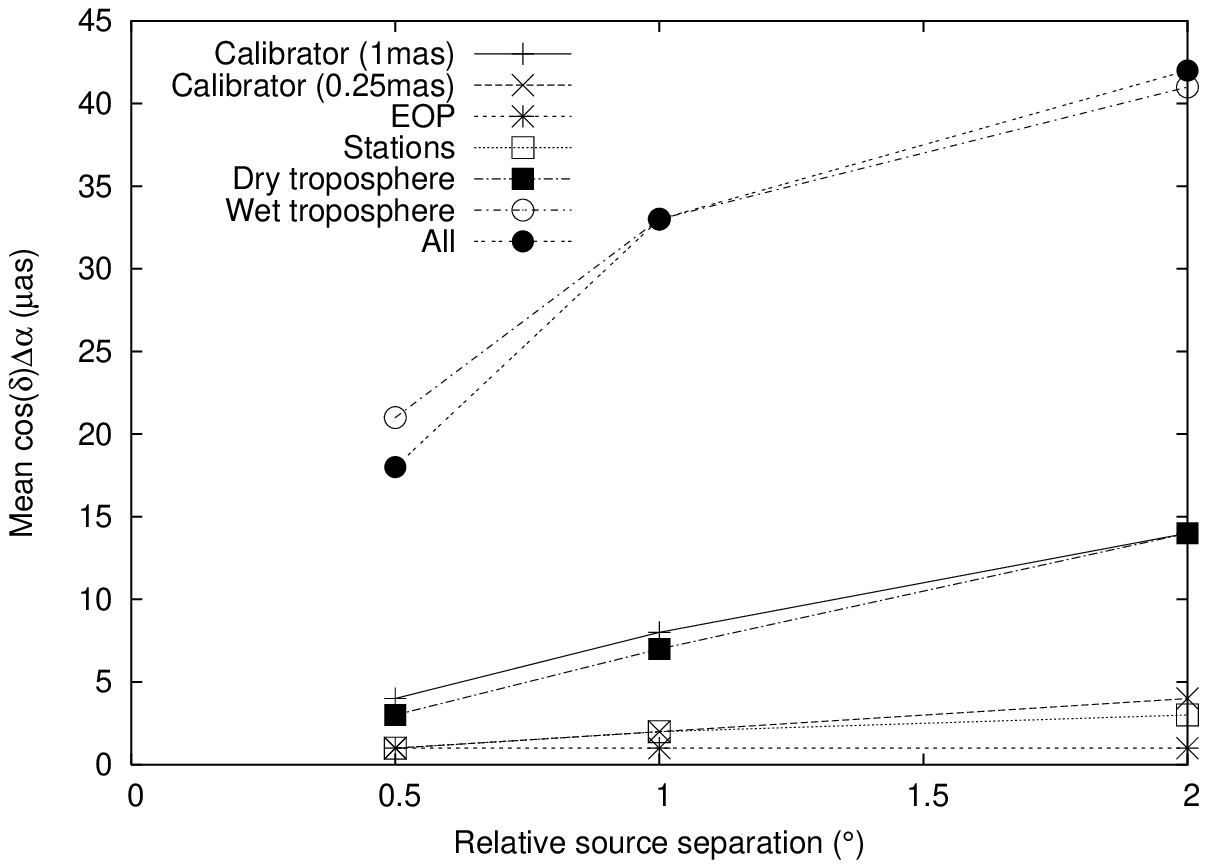}
\includegraphics[width=8cm,totalheight=8cm,angle=90, origin=c,angle=-90]{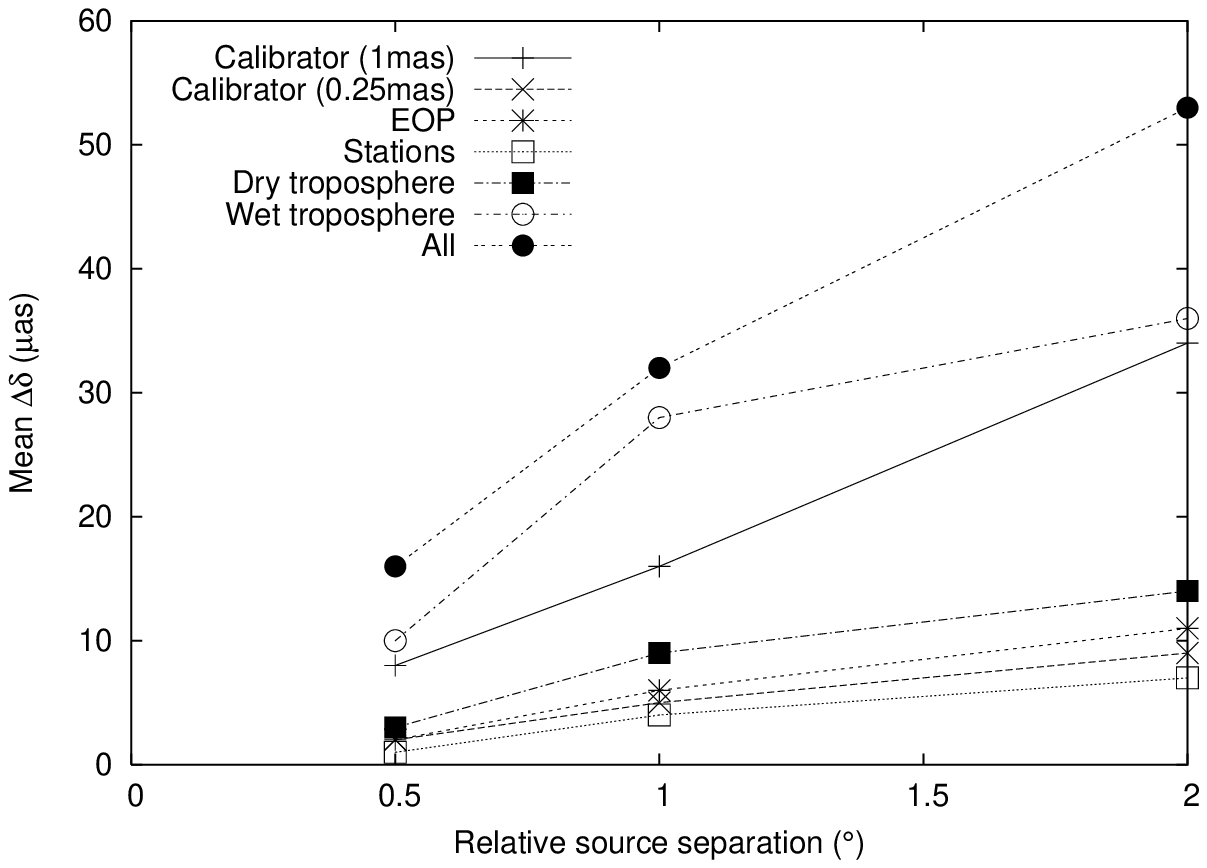}
\caption{Astrometric accuracy as a function of the relative source
separation for a target-calibrator pair observed with the VLBA at a
declination of $+25\degr$. Each error component is represented with a
different symbol and the total contributing error is also plotted.}
\label{plotseparation}
\end{figure*}

An important question is whether the astrometric accuracy scales linearly
as a function of the source separation. To study this matter, we
repeated all the previous simulations but with source
separations of $0.5\degr$ and $2\degr$. Then, we performed a linear fit
to the astrometric errors for the three values of the calibrator-target
separation ($0.5\degr$, $1\degr$ and $2\degr$), considering separately
each systematic error component of the tables above.
Figure~\ref{plotseparation} shows an example of such results for the
VLBA in the case of a target at $+25^\circ$ declination. Overall, our
plots show that the astrometric accuracy generally scales fairly linearly
as a function of the source separation.

To obtain a quantitative measure of the likehood of the
linearity, we determined the regression coefficients for each of the
107 linear fits. Such coefficients should be close to 1 for a linear
behavior while they should decrease as the behavior becomes less
linear. This analysis reveals that 80\% of the coefficients are larger
than $0.95$, indicating that the astrometric errors behave
linearly. Among all errors, calibrator position systematics are those
that were found to behave the least linearly. An empirical formula for
the astrometric accuracy $\Delta _{\alpha\cos\delta ,\delta}$ has been further
estimated by averaging the parameters of all the fits : 
\begin{equation}
\label{regression}
\Delta _{\alpha\cos\delta ,\delta} = (\Delta _{\alpha\cos\delta ,
\delta}^{1^{\circ}}-14) \times d+ 14~~~~~~(\mu as)
\end{equation}
where $\Delta _{\alpha\cos\delta ,\delta}^{1^{\circ}}$ is the astrometric error for $1\degr$
source separation as provided by our tables (\ref{VLBARA} and \ref{VLBADC}
for the VLBA, \ref{EVNRA} and \ref{EVNDC} for the EVN, \ref{GlobalRA} and
\ref{GlobalDC} for the global VLBI array) and $d=\sqrt{((\alpha -\alpha _0)\cos\delta_0)^2+(\delta-
\delta_0)^2}$ is the source separation in
degrees. In Section \ref{EVN}, we noted that the astrometric
accuracies of the EVN and the VLBA are similar, hence this formula should
apply to the EVN, too.

As a verification of this empirical formula, we computed the
astrometric accuracy for eight target-calibrator pairs observed
with the global VLBI array as part of a project to monitor
absolute lobe motions in compact symmetric objects
\citep{Cha03}. For the source pair
\object{J2212+0152}/\object{J2217+0220} with a separation of
$1.37\degr$ along the right ascension direction, we obtained simulated
accuracies $\Delta \alpha \cos\delta_0=42~\mu$as and $\Delta \delta
=63~\mu$as, versus $\Delta \alpha \cos \delta_0 =44~\mu$as and $\Delta
\delta  =63~\mu$as when derived from Eq. \ref{regression} and Table
\ref{GlobalRA}. In the worst case (target-calibrator
\object{J0754+5324}/\object{J0753+5352}  with a separation of
$0.50\degr$ along declination), simulated accuracies were $\Delta
\alpha \cos \delta_0=18~\mu$as and $\Delta \delta =12~\mu$as  while
Eq. \ref{regression} and Table \ref{GlobalDC} give $\Delta \alpha \cos
\delta_0=20~\mu$as and $\Delta \delta =24~\mu$as. Thus, overall we found
a discrepancy of a factor of 2 at most between our simple formula
(Eq.~\ref{regression}) and real simulation of the case considered.

\section{Conclusion}
\label{conclusion}

We have performed extensive simulations of VLBI data with the VLBA,
EVN and global VLBI array to study the dependence of the astrometric
accuracy on systematic errors in the phase model of phase-referenced
VLBI observations. Systematic errors considered in this study are
calibrator position uncertainties, station coordinate uncertainties,
Earth Orientation Parameters uncertainties and dry and wet troposphere
errors. We have adopted state of the art VLBI values for these errors.

Our simulations show that the astrometric accuracy of a full track
phase-referenced VLBI experiment is $50~\mu$as at mid declination and is
$\sim 300~\mu$as at low ($-25\degr$) and high ($+85\degr$)
declinations for point sources angularly separated by $1\degr$. Not
surprinsingly, the major systematic error originates from wet
tropospheric zenith delay uncertainties except at high declination where calibrator
position uncertainties dominate. We show that the astrometric
accuracy $\Delta _{\alpha\cos\delta ,\delta }$ depends linearly on the source 
separation and we established the simple formula $\Delta _{\alpha\cos\delta ,\delta }
= (\Delta _{\alpha\cos\delta , \delta}^{1^{\circ}}-14) \times d+ 14~~(\mu as)$
where $\Delta _{\alpha\cos\delta ,\delta}^{1^{\circ}}$ is
the astrometric error provided by our tables for the various arrays
and configurations and $d=\sqrt{((\alpha -\alpha _0)\cos\delta)^2+(\delta-
\delta_0)^2}$ is the source separation in degrees. Our
study has been carried out for point sources but  variable source structure
is likely to degrade the accuracy derived from this formula.

\appendix
\section{Analytical Behavior}

The analytical formulae in the Appendix A of \citet{Sha79} provide
the astrometric errors caused by the inaccuracy of the calibrator
coordinates in the case of a single VLBI baseline. Adopting our
notation, these formulae become~:
\begin{eqnarray*}
\Delta\alpha &\simeq &((\alpha-\alpha_0)\rm\tan\delta)\Delta\delta_0\\ &&-((\delta-\delta_0)\rm \tan\delta
+1/2\times(\alpha-\alpha_0)^2 ) \Delta\alpha_0 ,
\end{eqnarray*}
and
\begin{eqnarray*}
\Delta \delta &\simeq &(-(\delta -\delta_0) \cot \delta +1/2 \times
(\alpha -\alpha_0)^2)\Delta\delta_0 +\\
&&((\alpha-\alpha_0) \cot \delta)\Delta\alpha_0.
\end{eqnarray*}
where $\Delta\alpha$ and $\Delta\delta$ are the errors in
right ascension and declination introduced by errors $\Delta\alpha_0$ and $\Delta\delta_0$
in the coordinates of the reference source. The expression above for
$\Delta\delta$ restores correctly the last term of the equation which
was misprinted in the original paper.  These simple formulae are,
however, valid only for the special geometry adopted by the
authors where the ``baseline declination'' is $0^{\degr}$.

\begin{figure}[htb]
\centering
\includegraphics[width=8.5cm,totalheight=8.5cm]{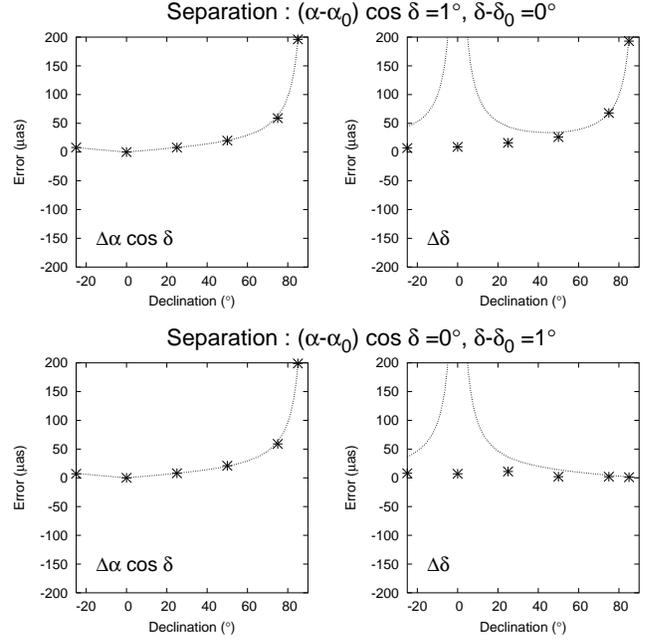}
\caption{Astrometric errors $\Delta\alpha\cos\delta$ and $\Delta\delta$
(respectively left and right) as a function of declination. The two upper
plots are for the case $\alpha-\alpha_0=(1/\cos\delta)\degr$ and $\delta -\delta_0 =0\degr$
while the two lower plots are for the case $\alpha -\alpha_0 = 0\degr$ and
$\delta -\delta_0 = 1\degr$. The continuous dotted lines show the errors
derived from the \cite{Sha79} formulae. The stars show the errors
from our simulations at six declinations from $-25\degr$ to $85\degr$.}
\label{shap}
\end{figure}

Adopting the same parameters as in our simulations ($\Delta\alpha_0= 1/\cos\delta_0$~mas,
$\Delta\delta_0=1$~mas, $\alpha -\alpha_0= 0\degr$ or $(1/\cos\delta_0)\degr$,
$\delta -\delta_0=1\degr$ or $0\degr$), we obtain the astrometric errors
plotted as a function of declination in Fig~\ref{shap} (dotted lines).
The results of our simulations for declinations of $-25\degr$, $0\degr$,
$25\degr$, $50\degr$, $75\degr$ and $85\degr$ in the case of the VLBA
(first lines of Tables 3 and 4) are also superimposed on these plots.

The right ascension errors obtained from the
simulations match perfectly those derived analytically, while the
declination errors show a strong discrepancy near declination
$0\degr$ (although they agree at high declinations). This discrepancy
originates from a singularity in the $\Delta\delta$ formula at
$\delta =0\degr$ (term in $\cot\delta$), inherent to the approximation
used to establish the formula (baseline declination of $0\degr$). For
a more complex and realistic network, such a singularity does not
exist, as also demonstrated by the results of our simulations.

\bibliography{3021}

\end{document}